\documentclass[prl, twocolumn, showpacs, amsmath]{revtex4}
\usepackage{dcolumn}% Align table columns on decimal point
\usepackage{bm}% bold math
\usepackage{graphicx}% Include figure files
\usepackage{color}

\DeclareGraphicsExtensions{. jpg,. pdf, . mps, . png, . eps, . ps, . EPS}

\begin{document}
\def\be{\begin{equation}}
\def\ee{\end{equation}}

\def\bc{\begin{center}} 
\def\ec{\end{center}}
\def\bea{\begin{eqnarray}}
\def\eea{\end{eqnarray}}
\newcommand{\avg}[1]{\langle{#1}\rangle}
\newcommand{\Avg}[1]{\left\langle{#1}\right\rangle}
\newcommand{\lap}[1]{\nabla^{#1}}
 
\title{Strong Noise Effects in one-dimensional Neutral Populations}

\author{Luca Dall'Asta$^1$, Fabio Caccioli$^2$, and Deborah Begh\`e$^3$}

\affiliation{$^1$ Abdus Salam International Center for Theoretical 
Physics, Strada Costiera 11, 34151 Trieste, Italy\\
$^2$ Santa Fe Institute, 1399 Hyde Park Road, Santa Fe, NM 87501, USA\\
$^3$ Dipartimento di Biologia Evolutiva e Funzionale, Universit\`a degli Studi di Parma,
Viale G.P. Usberti 11/A, I-43124 Parma, Italy\\ 
}

\begin{abstract}
The dynamics of well-mixed biological populations is usually studied by mean-field methods and weak-noise expansions. Similar methods have been applied also in spatially extended problems, relying on the fact that these populations are organized in colonies with a large local density of individuals. 
We provide a counterexample discussing a one-dimensional neutral population with negative frequency-dependent selection. The system exhibits a continuous phase transition between genetic fixation and coexistence unexpected from weak-noise arguments. We show that the behavior is a non-perturbative effect of the internal noise that is amplified by presence of spatial correlations ({\em strong-noise regime}). 
\end{abstract}
%\pacs{89. 75-k, 89. 75. Fb, 89. 75. Hc}

\maketitle
The dynamics of biological populations is traditionally modeled using systems of partial differential equations for a set of density fields, representing the (local) concentrations of species, resources, genetic traits, etc. \cite{M03}. Despite the success of deterministic theories,
% such as  Lotka-Volterra equations for species competition \cite{LV20} and Fisher's equation for the diffusion of a beneficial mutation in a genetic population \cite{FKPP37}, 
the fate of real populations is often determined by purely stochastic forces induced by the discreteness of individuals ({\em internal noise}). 
In well-mixed populations, the internal noise is a finite-size effect that can be studied using system's size expansions \cite{VK92}. Even though the noise has macroscopic effects, including gigantic oscillations and unexpected extinction events \cite{MN05}, these results are obtained in a weak-noise regime accessible from the deterministic mean-field description by perturbative methods. 
In spatially extended populations, on the contrary, correlations induced by spatial constraints can enhance the effects of the internal noise and induce phenomena that are completely unpredictable by means of perturbative methods. 
Spatial populations are usually studied using ``patch dynamics" \cite{DL98} or ``stepping stone models" \cite{KW64}, in which colonies composed of a number ($\Omega \gg 1$) of individuals occupy the sites of a lattice and interact by migration.
All processes inside the colonies (or {\em demes}) and between them are stochastic, therefore as long as their internal population is finite and mobility is low, the macroscopic behavior is affected by noise. Unlike the well-mixed case, these phenomena can persist also in the limit of infinitely large systems. 
The simplest situation occurs in {\em neutral theories} \cite{CK70}, in which purely stochastic  genetic (ecological) drift is the only force determining the evolution of the whole population, favoring local exclusion and fixation. In low dimensional systems, genetic drift competes with diffusion and migration processes that instead promote genetic mixing \cite{KW64}. 
When the microscopic interactions are linear, like in the voter model or in the Moran process, changing microscopic details of the model (e.g. $\Omega$) only produces a temporal rescale of the dynamics. In the presence of non-linear interactions, instead, the continuum theory may depend on the microscopic model and on the way in which the infinite size limit is taken.

In this Letter we address this problem considering the effects of  {\em negative frequency-dependent selection} (NFS) in a neutral population composed of two competing genetic types.
In a regime of strong competition, individuals carrying a rare phenotype are able to exploit new resources, survive predation or better adapt to an evolving environment, gaining a fitness advantage. 
This advantage is transient, because it declines when the phenotype becomes more common.  
In this sense, it is a special form of Darwinian selection that, within the framework of neutral theories, favors advantageous mutations and {\em maintains balanced polymorphism and biodiversity}. 
Negative frequency-dependent selection has been recently observed experimentally in natural populations both at the phenotype and genetic level \cite{FFRS07}.
While NFS is enough to maintain a stable coexistence of different genetic types in well-mixed populations, we show that in one-dimensional systems the competition between NFS and genetic drift leads to an unexpected continuous phase transition between genetic fixation and coexistence. 

We consider an idealized population of identical haploid individuals that differ in a single gene, i.e. at a given locus they have either allele A or B.  According to Kimura's stepping stone models \cite{KW64}, the population is spatially organized in ``demes", i.e. islands or colonies that occupy the sites of a $d$-dimensional lattice. Each deme contains a fixed number $\Omega$ of individuals.  The evolution of the population is governed by three stochastic processes: 1) a {\em migration process} (at rate $\lambda_D$) that consists  in the pair-exchange of  individuals between neighboring demes; 
2) a {\em Moran process}  (at rate $\lambda_M$)  in which an individual (either of type A or B)  is chosen for reproduction and her offspring replaces another random individual of the same deme; 3) a process of {\em  frequency-dependent selection} at rate $\lambda_S$.  The latter consists in randomly choosing an individual together with other two individuals of the same deme. The first individual is replaced by a new one carrying the less/most popular allele appearing in the triple. In the case of NFS, less frequent alleles are selected. 
%This process mimics competition for reproduction and selection based on a frequency-dependent principle. 
%For instance, If allele A appears once and B twice, the first will be selected for replacement.  Instead nothing will happen if all three individuals carry allele A (or B). 
%The processes conserve the size of the demes, that is fixed to $\Omega$.
%Note that the theory is {\em neutral}, that is alleles A and B are equally fit and there is no neat selection advantage for one of them, only a local advantage in favor of rare traits.   
The state of the system is univocally described by a vector $\{n_1, n_2, \dots, n_{L^d}\} \equiv \underline{n}$ where $n_i$ is the number of individuals of type A in deme $i$ and $L$ is the linear size of the lattice. The evolution of this state is governed by the master equation $d P(\underline{n},t)/dt = \sum_{\underline{n}' \neq \underline{n}} \left[ W(\underline{n}' \to \underline{n}) P(\underline{n}',t) - W(\underline{n} \to \underline{n}') P(\underline{n},t) \right]$, with local transition rates 
\begin{subequations}
\begin{eqnarray} 
w_D(n_i \to n_i-1,n_j \to n_j+1) = & \frac{\lambda_{D}}{2d}  \frac{n_i}{\Omega} \frac{\Omega-n_j}{\Omega}\\
w_M(n_i \to n_i-1) = & \lambda_{M}  \frac{n_i}{\Omega} \frac{\Omega-n_i}{\Omega} \\ 
w_S(n_i \to n_i-1) = & \lambda_S \left(\frac{n_i}{\Omega}\right)^2   \frac{\Omega-n_i}{\Omega}.
\end{eqnarray}
\end{subequations}
Following standard stochastic methods \cite{VK92,MN05}, we write $n_i = \Omega f_i + \mathcal{O}(\Omega^\frac{1}{2})$, and exploiting the fact that frequencies change of $\mathcal{O}({\Omega^{-1}})$ in a time step $\Delta t$, we expand the master equation  in powers of $1/\Omega$. In the limit of large $\Omega$, after temporal rescaling $t = \Omega \Delta t$, we obtain a Fokker-Planck equation and the corresponding non-linear Langevin equation for the local frequency $f_i$ of type A. We take a naive continuum limit sending the lattice spacing $a$ to zero and keeping constant the diffusion coefficient $D = a^2 \lambda_D$. The effective continuum equation for $f(x,t)$ reads
\begin{equation}
\label{omegalarge}
\frac{d f}{dt}  = D  \lap{2}f - \nu (2 f-1) f(1-f) +  \sigma \sqrt{f(1-f)} \eta 
\end{equation}
where $\nu = \lambda_{S}$, $\sigma^2 = (\lambda_M+\lambda_{S})/\Omega$ and $\eta(x,t)$ is a delta correlated gaussian noise ($\langle \eta(x,t)\eta(x',t') \rangle =  \delta(t-t')\delta(x-x')$).  
\begin{figure}
\includegraphics[width=0.8\columnwidth]{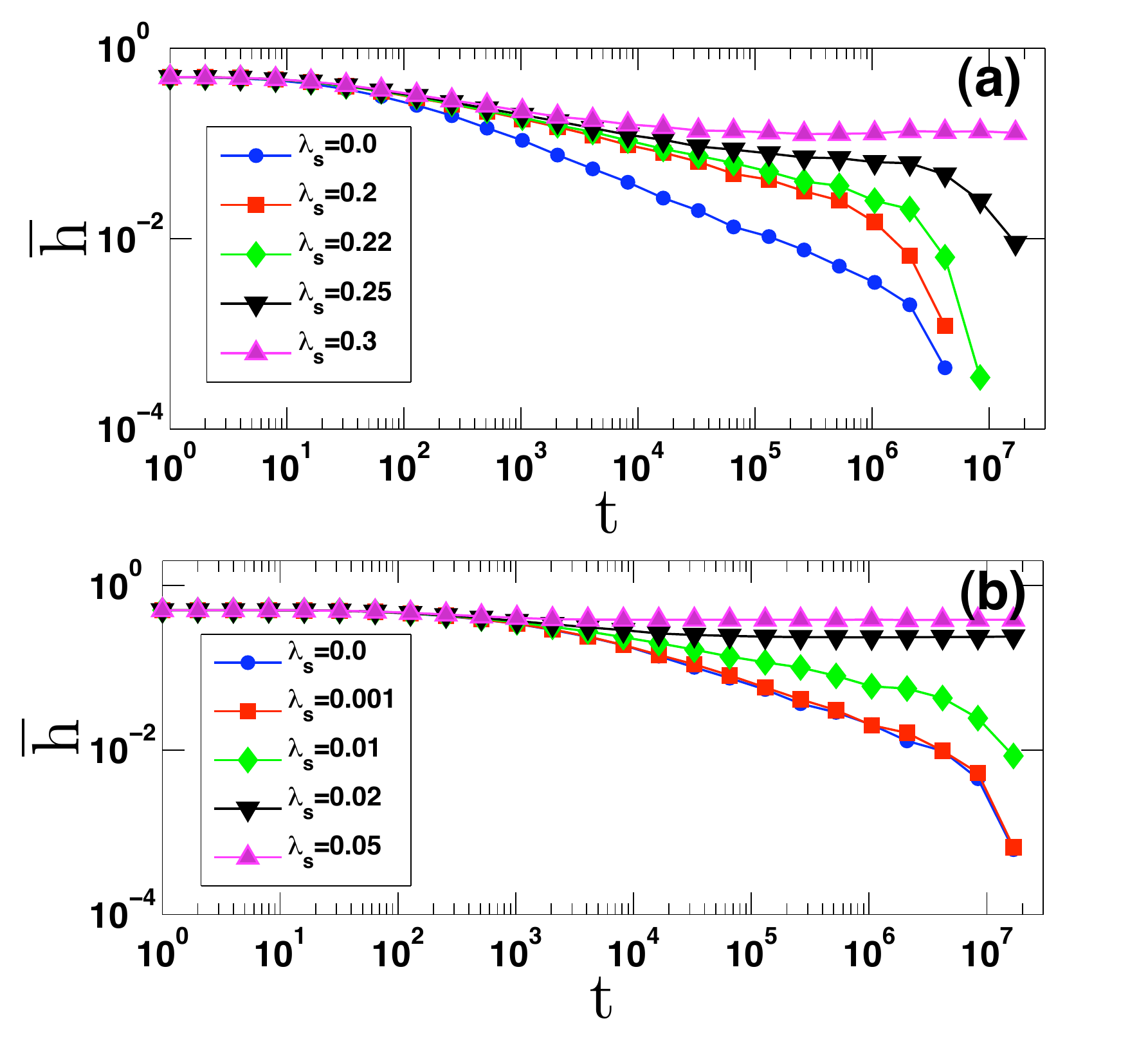}
\caption{Time dependence of the average heterozygosity $\bar{h}(t)$ for a system of size $L=10^3$, $\lambda_D=0.2$, $\lambda_M=1$, and different values of $\lambda_S$. (a) $\Omega = 5$ and $\lambda_S = 0, 0.2, 0.22, 0.25, 0.3$. (b) $\Omega = 20$ and $\lambda_S = 0, 0.001, 0.01, 0.02, 0.05$. Numerical data are averaged over $100$ initial conditions.
}\label{fig1}
\end{figure}
The parameter $\nu$ can be made negative admitting also positive frequency-dependent selection. In this case, the two symmetric absorbing states at $f=0,1$ are attractive and the system fixates to one of them after a coarsening process. 
For $\nu > 0$ instead, the attractive fixed point is the active homogeneous equilibrium state at $f_{eq}=1/2$, corresponding to a completely mixed {\em coexistence}. The fluctuations around $f_{eq}$ are gaussian with amplitude $\mathcal{O}(\Omega^{-1/2})$.
The mean-field picture, with a discontinuous phase transition at $\nu_c=0$, is not correct in one dimension.
This is evident from the results of numerical simulations  in Fig.\ref{fig1}, where we plot the average heterozygosity $\bar{h}(t) = L^{-1}\int_0^L \langle f(x,t)(1-f(x,t)) \rangle dx$ as a function of time for a system of size $L=10^3$ (with $\lambda_D = 0.2$, $\lambda_M=1$) and different values of $\lambda_S$ and $\Omega$. In a finite system, the dynamics cannot escape the absorbing state and $\bar{h}$  eventually vanishes. However, the temporal behavior of $\bar{h}$ clearly shows the existence of a phase transition between fixation and coexistence at some non-zero $\lambda_S$ (i.e. $\nu_c >0$). For $\Omega =5$ the coexistence phase develops gradually, indicating the continuous nature of the transition. Increasing $\Omega$ (Fig.\ref{fig1}b), the critical point moves closer to zero, and the transition becomes more abrupt. In this case, numerical results do not allow to distinguish  between continuous or discontinuous behavior.
A discontinuous transition is recovered for sure in the thermodynamic limit, at least if we first take $\Omega \to \infty$ and then $L\to \infty$ limits. 
An argument in favor of the continuous character of the transition is based on Hinrichsen's conjecture about the impossibility of first-order transitions in non-equilibrium $1d$ systems with short-range interactions \cite{H00}. Indeed, in the absence of surface tension ($d=1$), a small droplet of the absorbing phase can always grow by diffusion of the boundaries and eventually destabilize the fluctuating active equilibrium. For $\Omega<\infty$, the nucleation rate $R_\ell$ of a droplet of size $\ell$ of the absorbing phase out of the homogeneous equilibrium is small but finite ($R_\ell \approx e^{-c \Omega\ell}$ with $c \propto \frac{\lambda_S}{\lambda_M+\lambda_S}$). Hence, if the only processes  into play were nucleation and diffusion,  the absorbing phase would always dominate. It is thus necessary to have a  different mechanism for the emergence of coexistence. 
 According to this argument, the Fokker-Planck approach and small-noise expansions are not valid even in the large $\Omega$ limit, calling for a different ``non-perturbative approach".
Interestingly, a non-perturbative renormalization group study of Eq.\ref{omegalarge} was done in \cite{CCDDM05} showing that, for $d< d'_c \approx 4/3$, the critical behavior of the model is controlled by a non-Gaussian fixed point of the RG flow. In fact,  \cite{CCDDM05} suggests the existence of a continuous phase transition at $\nu_c >0$ that cannot be found by expanding in $1/\Omega$ around the mean-field theory.

A way to confirm and understand this picture is that of studying a model with only one individual per site and interactions with neighboring sites.  
The migration process is unchanged and consists of exchanging two neighboring individuals with rate $\lambda_D$. The Moran process is reduced to a voter-like move between nearest neighbors. With rate $\lambda_M$, an individual dies and is immediately replaced by an offspring of a randomly selected neighbor.
We adopt a spatial version of NFS  recently proposed in theoretical ecology to model interspecies competition \cite{NP99}.  Two neighboring individuals that are different compete for reproduction: with a rate $\lambda_{S}$ we randomly select a neighbor of the two to be replaced with the less common type among the three selected individuals. 
In the absence of diffusion ($\lambda_D=0$), the one-dimensional dynamics can be mapped on that of  branching-annihilation random walks with even offsprings (BARWe) \cite{CT96} or it can be analyzed with approximated techniques developed for binary spin systems. After the mapping $\sigma_i=2n_i-1$, the transition rates take the simple form $w_M(\sigma_i \to -\sigma_i)$ =  $\frac{\lambda_{M}}{2} \frac{1-\sigma_i \sigma_{i\pm1}}{2}$,  $w_{S}(\sigma_i \to -\sigma_i)$  =  $\frac{\lambda_{S}}{2} \frac{1-\sigma_{i\pm1}\sigma_{i\pm2}}{2}$.
From the master equation for $P(\underline{\sigma},t)$, we can compute coupled evolution equations for the multi-spin correlation functions $\langle \sigma_i \dots \sigma_j\rangle$  that cannot be solved exactly. At the lowest order of the hierarchy we have 
\begin{widetext}
\begin{eqnarray}
\frac{d \langle \sigma_i \rangle}{dt} & = & \frac{\lambda_M}{2} \left[ \langle \sigma_{i-1} \rangle + \langle \sigma_{i+1} \rangle - 2 \langle \sigma_i \rangle \right] + \frac{\lambda_S}{2} \left[ \langle \sigma_{i} \sigma_{i-1} \sigma_{i-2} \rangle + \langle \sigma_{i} \sigma_{i+1} \sigma_{i+2}\rangle - 2 \langle \sigma_i \rangle \right] \\  
\nonumber \frac{d \langle \sigma_i \sigma_{j} \rangle}{dt} & = & \frac{\lambda_M}{2} \left[ \langle \sigma_i (\sigma_{j+1} + \sigma_{j-1}) \rangle  +  \langle \sigma_j (\sigma_{i+1} +\sigma_{i-1}) \rangle  -4 \langle \sigma_{i}\sigma_{j} \rangle \right]  \\
 & & +  \frac{\lambda_S}{2}  \left[ \langle  \sigma_{i} \sigma_{j} (\sigma_{i-1} \sigma_{i-2} + \sigma_{i+1} \sigma_{i+2})\rangle +  \langle  \sigma_{i} \sigma_{j} (\sigma_{j-1} \sigma_{j-2} + \sigma_{j+1} \sigma_{j+2})\rangle  - 4\langle \sigma_i \sigma_{j}  \rangle \right] .
\end{eqnarray}
\end{widetext}
Some progress can be done by assuming the Kirkwood approximation, i.e. factorizing three-sites and four-sites quantites as products of lower order ones \cite{LR07}. For convenience we define  the average frequency of type $A$ as $\bar{f}_i=(1+\langle \sigma_i\rangle)/2$ and the average local heterozygosity  $\bar{h}_i =(1-\langle \sigma_i \sigma_{i+1} \rangle)/2$.  Taking the continuum limit, and neglecting higher order contributions, we get equations for $\bar{f}(x,t)$ and $\bar{h}(x,t)$,
\begin{eqnarray}
\frac{\partial \bar{f}}{\partial t} & = & D_M \frac{\partial^2 \bar{f}}{\partial x^2} - \lambda_S (2 \bar{f} - 1)\bar{h} \label{eqsferm1}\\
 \frac{\partial \bar{h}}{\partial t} & = & (D_M + D_S)  \frac{\partial^2 \bar{h}}{\partial x^2}  - 2 \bar{h}\left(\lambda_M - \lambda_S (1 -2\bar{h})\right) \label{eqsferm2}
\end{eqnarray}
where $D_M = a^2 \lambda_M$ and $D_S(\bar{h}) = a^2 \lambda_S \bar{h}(x,t)$.
The effective diffusion terms generated in the continuum limit account for the random walk of domain walls induced by both Moran and NFS moves. The coefficient $D_S(\bar{h})$ depends linearly on the local heterozygosity $\bar{h}(x,t)$ as required by NFS microscopic rule.    
Eqs. \ref{eqsferm1}-\ref{eqsferm2} are deterministic and describe the average behavior of the system. A single realization of the process would also present a  noise term with two homogeneous absorbing states at $f(x,t) = 0,1$ $\forall x$. 
\begin{figure}[b]
\includegraphics[width=1.\columnwidth]{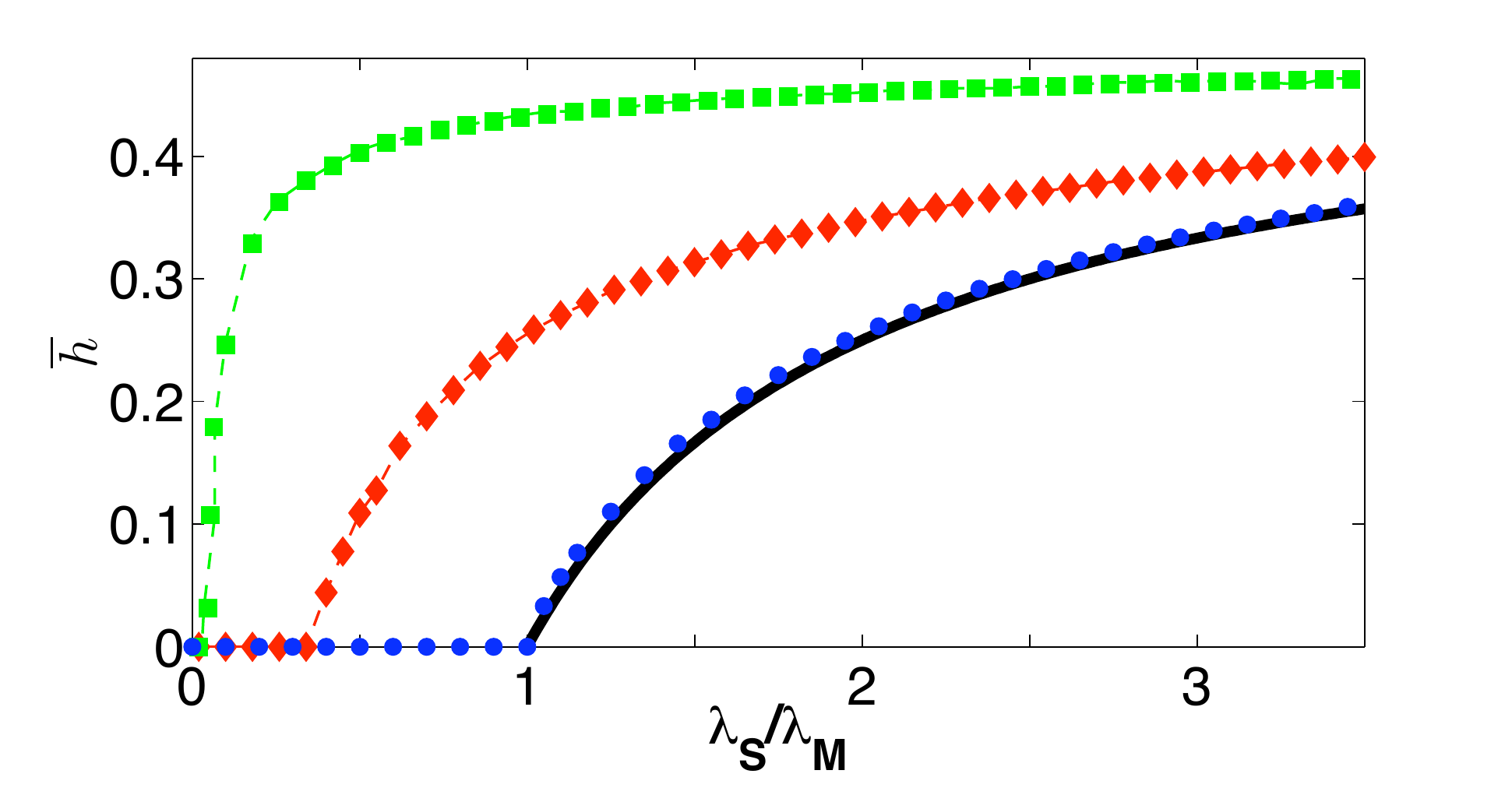}
\caption{Average heterozygosity $\bar{h}$ in the stationary state as a function of the control parameter $\lambda_S$ (with $\lambda_M=1$) for the model with single individual per site. Numerical simulations (points) are performed on a system of size $L=10^3$ for different migration rates: $\lambda_D=0$ (blu circles), $2$ (red diamonds), $20$ (green squares). The case $\lambda_D=0$ is in good agreement with the stationary solution of Eqs. \ref{eqsferm1}-\ref{eqsferm2} (full line).
}\label{fig2}
\end{figure}
The stationary state is obtained solving Eqs. \ref{eqsferm1}-\ref{eqsferm2} with the assumption of spatial translation invariance, $\bar{f}(x) \approx \bar{f}$ and $\bar{h}(x) \approx \bar{h}$. 
There are two stationary solutions of Eq.\ref{eqsferm1} depending on the ratio $\lambda_S/\lambda_M$. A continuous phase transition separates the absorbing phase  ($\bar{h}=0$) where a unique species survives from an active phase of non-trivial coexistence ($\bar{h}=\frac{\lambda - 1}{2 \lambda}$ with $\lambda = \lambda_S/\lambda_M$). 
The theoretical curve $\bar{h}(\lambda_S/\lambda_M)$ is plotted in Fig.\ref{fig2} (full line) together with the results of numerical simulations (points). For $\lambda_D=0$ the agreement is very good, although in the analytical calculation we neglected the effects of the self-generated spatial diffusion that could be responsible of the renormalization of both the critical point and the critical exponents. 
To explore the dynamics in the absorbing phase, we prepared the population of size $L=10^4$ in a completely mixed configuration ($\bar{f}=0$, $\bar{h}=1/2$) and we studied the temporal behavior of the average cluster size (of type A or B) until fixation. Clusters growth is evident from the time-oriented snapshots in Fig. \ref{fig3}a. The average domain length is roughly $\ell(t) \propto 1/\bar{h}(t)$, therefore the behavior Eq. \ref{eqsferm2} provides information on the coarsening exponent. In the absorbing region, we can neglect the non-linear terms in the r.h.s. and solving the linear equation we obtain a power law decay $\bar{h}(t) \propto t^{-1/2}$. This is confirmed by the numerical results shown in Fig. \ref{fig3}b. The typical domain length $\ell(t)$ thus grows in time as $\ell(t) \sim t^{1/2}$, which is the same behavior as the pure voter model (i.e. $\lambda_S=0$). This law is not modified by small diffusion, but the case $\Omega>1$ requires further investigation (see Fig.\ref{fig1}).
In the active phase, the system shows local clustering up to a characteristic length scale $\ell \propto 1/\bar{h}$, that is reduced by increasing $\lambda_S$ from just above criticality (Fig.\ref{fig3}c) to the bulk of the coexistence region (Fig.\ref{fig3}d). Comparing snapshots in Figs.\ref{fig3}c-\ref{fig3}d with that in Fig.\ref{fig3}a, we notice a much higher local volatility of patches and domain walls, a signature of increasing genetic mixing. 
The addition of a small migration rate does not change the qualitative picture observed for $\lambda_D =0$. However, increasing diffusion favors local mixing of individuals and acts against genetic drift. 
This is confirmed by numerical simulations reported in Fig.\ref{fig2}, that show the critical point moving towards zero and the curve of $\bar{h}$ becoming steeper  for increasing values of $\lambda_D$.
In the limit $\lambda_D \to \infty$ the constraints imposed by the one-dimensional geometry are eliminated leaving a completely well-mixed population that should undergo a first-order transition at zero NFS as predicted by the mean-field theory. 

\begin{figure}
\includegraphics[width=4.25cm,height=3.15cm]{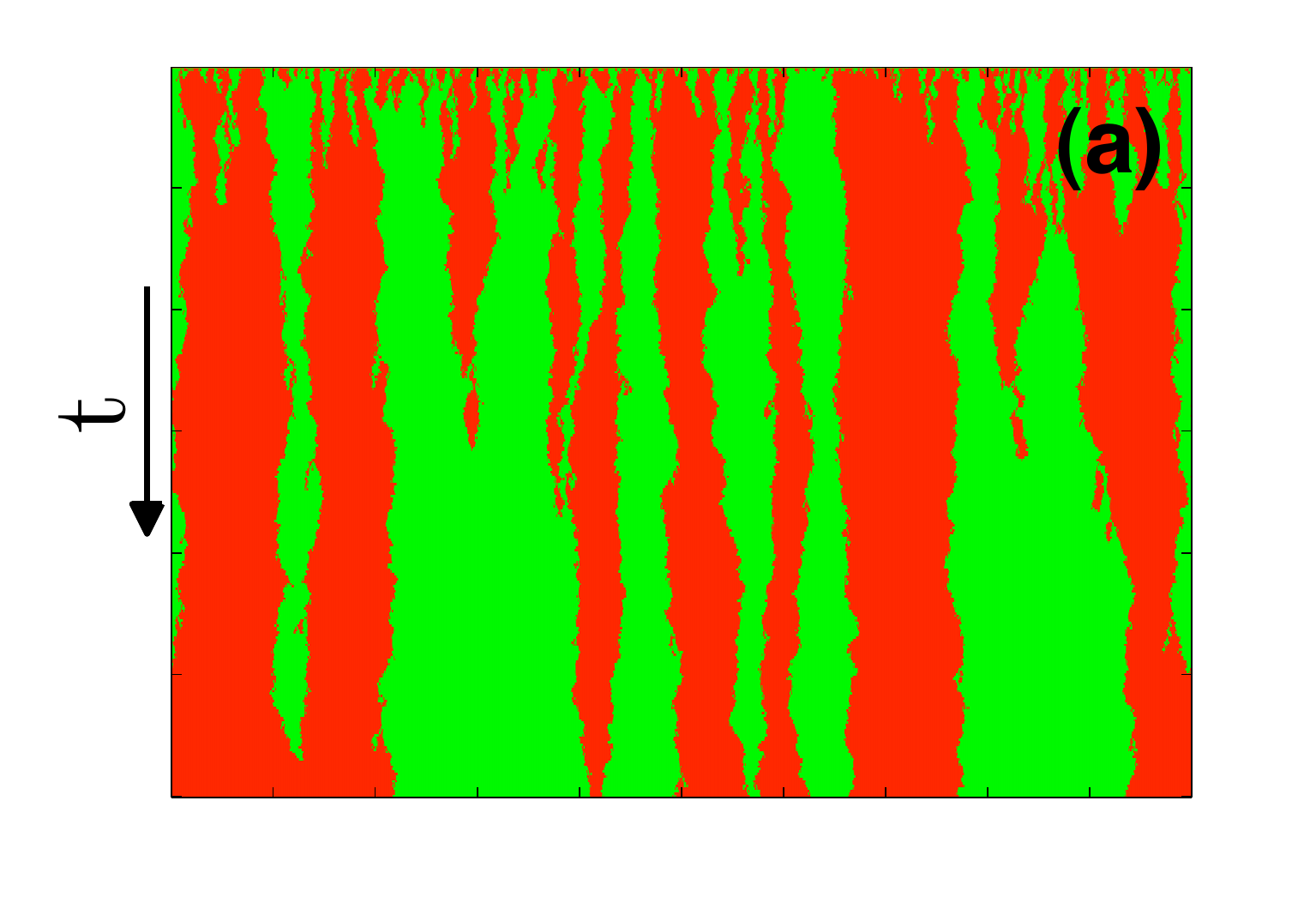}
\includegraphics[width=4.25cm,height=3.15cm]{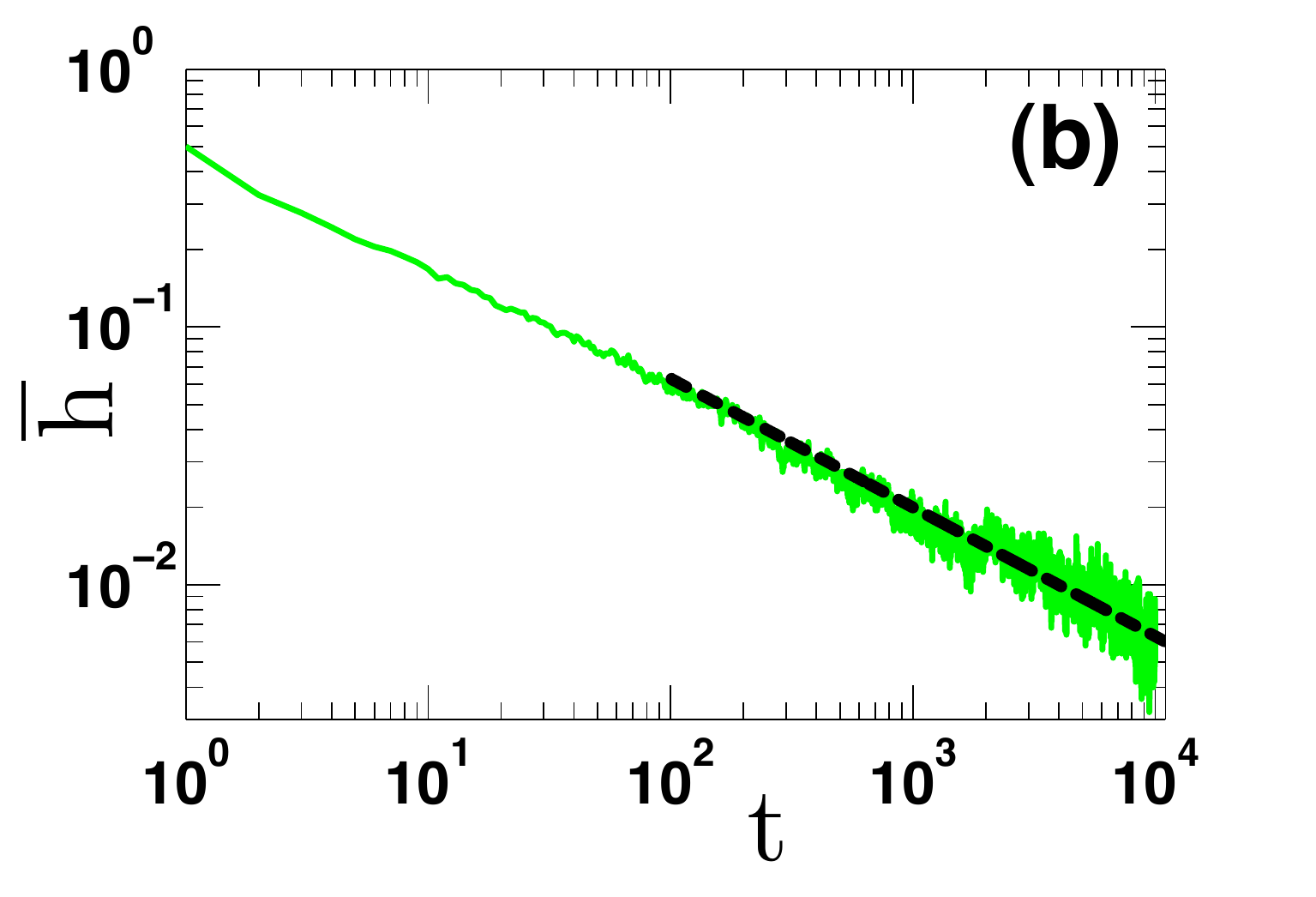}\\
\includegraphics[width=4.25cm,height=3.15cm]{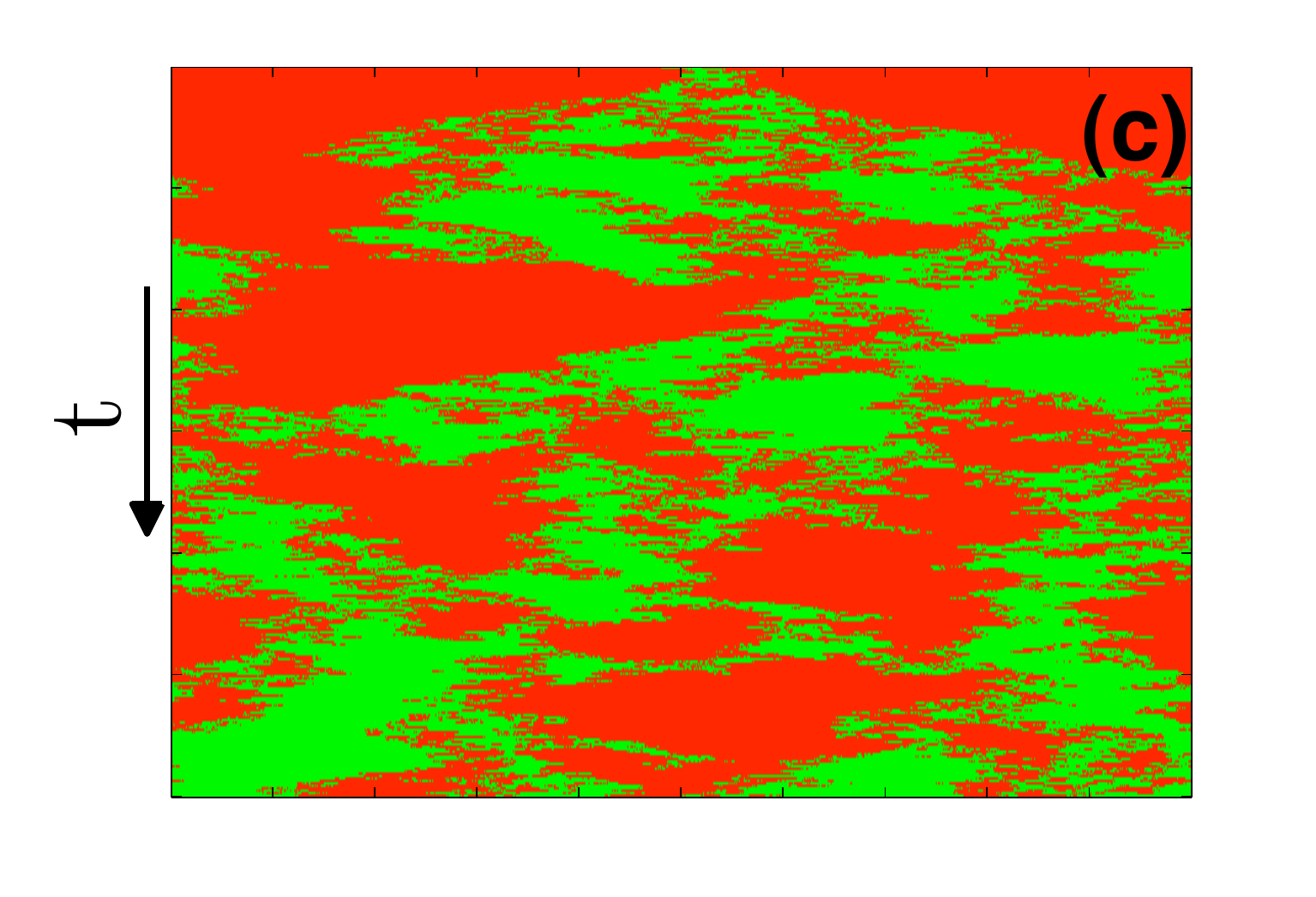}
\includegraphics[width=4.25cm,height=3.15cm]{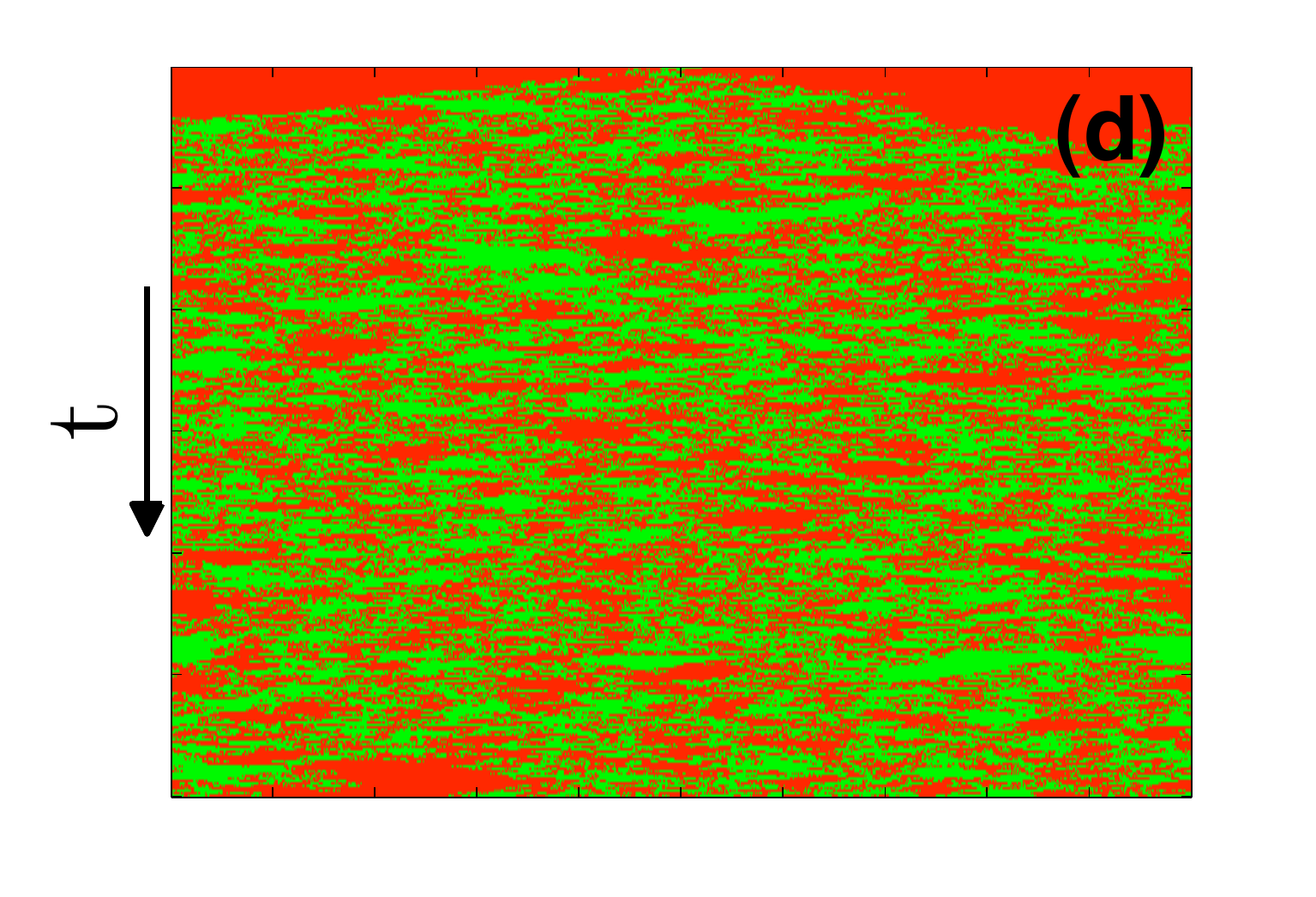}
\caption{(a) $1$+1 dimensional snapshot of the evolution of a system of size $L=10^4$ starting from random initial conditions into the absorbing phase ($\lambda_S=0.5$); (b) temporal decay of the average local heterozygosity $\bar{h}$ for $\lambda_S=0.5$, the dashed line indicates a power-law decay $\propto t^{-0.5}$;
(c-d) $1$+1 dimensional snapshots of the evolution of a system of size $L=10^4$ starting from a single defect towards the stationary active phase for $\lambda_S=1.05$ (c) and $1.5$ (d).
}\label{fig3}
\end{figure}

In conclusion, negative frequency-dependent selection is a major force for coexistence, but spatial constraints tend to amplify the effects of genetic drift promoting fixation. The outcome of this competition  is a phase transition between a monomorphic and a polymorphic phase. In one spatial dimension, the nature of the phase transition critically depends on the number $\Omega$ of individuals per lattice site and on their mobility $\lambda_D$.  In the infinite size limit, any finite value of $\Omega$ generates a dynamics that cannot be obtained from the mean-field by means of perturbative approaches ({\em strong noise regime}).  Recent works have shown that the strong noise limit can be important in the case of weak directional selection, therefore the present results could be applied to a large class of phenomena occurring at the one-dimensional  fronts of expanding populations, such as Fisher's waves \cite{HK09} and  ``gene surfing" \cite{HHRN07}.
Negative frequency-dependent selection is also expected to play a role in plant communities and marine ecosystems. It would be interesting to investigate whether one-dimensional structures such as valleys, riverbanks and seacoasts, could act as spatial bottlenecks promoting or impeding genetic variability even in higher dimensional populations.

The authors acknowledge A. C. Barato  for pointing out reference \cite{H00} and T. Galla for a critical reading of the manuscript.

\end{document}